\documentclass{llncs}

\usepackage{amssymb}
\usepackage{amsmath}
\usepackage{color}
\usepackage{graphics,latexsym,amsfonts}
\usepackage{graphicx}
\usepackage{eurosym}
\usepackage{subfigure}
\usepackage{multicol}
\usepackage{pstricks, pst-node}

\newcommand{\IM}{\mathbb{M}}
\newcommand{\IT}{\mathbb{T}}
\newcommand{\IU}{\mathbb{U}}
\newcommand{\IR}{\mathbb{R}}

\advance\textwidth by 2.5truecm
\advance\oddsidemargin by -1truecm
\advance\textheight by 5.5truecm \advance\topmargin by -2.5truecm

\begin{document}
\title{Global Green Economy and Environmental Sustainability:
a Coopetitive Model}

\author{David Carf\`\i*,  Daniele Schilir\`o**}

\institute{*Department of Mathematics\\
University of California at Riverside\\
900 Big Springs Road, Surge 231 Riverside, CA 92521-0135, USA.\\
**Department DESMaS, \\
University of Messina,\\
Via Tommaso Cannizzaro 275, Me 98122, Messina, Italy.\\
 \email{*davidcarfi71@yahoo.it}\\
 \email{**danieleschiliro@unime.it}}

\date{\today}

\maketitle

%

\begin{abstract}

\noindent

This paper provides a coopetitive model for a global green economy, taking into account the environmental sustainability. In particular, we propose a differentiable coopetitive game $G$ (in the sense recently introduced by D. Carf\`{\i}) to represent a global green economy interaction, among a country $c$ and the rest of the world $w$. Our game $G$ is a linear parametric (Euclidean) perturbation of the classic Cournot duopoly. In the paper we offer the complete study of the proposed model and in particular a deep examination of its possible coopetitive solutions.

\medskip


\end{abstract}


\section{Introduction}

A major driver of human impact on Earth's systems is the destruction of biophysical resources. The accelerating carbon emissions, in particular, indicate a mounting threat of climate change with potentially disastrous human consequences. The total environmental impact of a community or of humankind as a whole on the Earth's ecosystems depends on both population (which is increasing) and impact per person, which, in turn, depends in complex ways on what resources are being used, whether or not those resources are renewable, and the scale of the human activity relative to the carrying capacity of the ecosystems involved.  

\subsection{Environmental Sustainability (ES)}
   
Thus, we are facing a problem of environmental sustainability (ES). Goodland in \cite{Goodland} defines ES as the maintenance of natural capital. ES emphasizes the environmental life-support systems without which neither production nor humanity could exist. These life-support systems include atmosphere, water and soil; all of these need to be healthy, meaning that their environmental service capacity must be maintained. So ES can be represented by a set of constraints on the four major activities regulating the scale of the human economic subsystem: the use of renewable and nonrenewable resources on the source side, and the pollution and waste assimilation on the sink side. This is why ES implies that careful resource management be applied at many scales, from  economic sectors like agriculture, manufacturing and industry, to work organizations, to the consumption patterns of households and individuals and to the resource demands of individual goods and services.

\subsection{Geopolitical situation}

Currently, there is no international consensus on the problem of global food security or on possible solutions for how to nourish a population of 9 billion by 2050.  Freshwater scarcity is already a global problem, and forecasts suggest a growing gap by 2030 between annual freshwater demand and renewable supply \cite{United Nation}. The increasing dependence on fossil fuels, the problems of security of supply and the best solutions for mitigating climate change require important and urgent measures that must be adopted at a global level. The economic and geopolitical events are changing the international context and also the costs of energy supply in the world industrialized nations. The financial and economic crisis of 2008, in particular, has caused a widespread disillusionment with the prevailing economic paradigm, and it has determined a strong attraction for the green economy.

\subsection{Green Economy} The green economy is an economy concerned with being:

\begin{enumerate}

\item environmentally sustainable, based on the belief that our biosphere is a closed system with finite resources and a limited capacity for self-regulation and self-renewal. We depend on the earthÕs natural resources, and therefore we must create an economic system that respects the integrity of ecosystems and ensures the resilience of life supporting systems;

\item socially just, based on the belief that culture and human dignity are precious resources that, like our natural resources, require responsible stewardship to avoid their depletion;

\item locally rooted, based on the belief that an authentic connection to place is the essential pre-condition to sustainability and justice. The global green economy is an aggregate of individual communities meeting the needs of its citizens through the responsible production and exchange of goods and services. 

\end{enumerate}

Thus global green economy growth is compatible with environmental sustainability over time, since it seeks to maintain and restore natural capital. 

\subsection{The coopetition}

This paper applies the notion of coopetition within a macro-global context for environmental problems. The concept of coopetition is a complex construct originally devised by Brandenburger and Nalebuff (\cite{Branden1996}) in strategic management at microeconomic level (see \cite{Schi 2009}). According to this concept, the economic agents (i.e. firms or countries) must interact to find a win-win solution, that indicates a situation in which each agent thinks about both cooperative and competitive ways to change the game. So in our case, each country has to decide whether it wants to collaborate with the rest of the world in getting an efficient global green economy, even if the country is competing in the global scenario.

\subsection{Our model}

The coopetitive model here suggested is environmentally sustainable, since it should lead to \textit{maintain natural capital}, by using mainly renewable resources. This \textit{ES coopetitive model} also aims at reducing emissions of greenhouse gases, determining the reduction of global pollution, in this way it contributes to the establishment of a sustainable and lasting global green economy. Finally, the model determines a change in the patterns of consumption of households towards goods and human behaviors with a lower environmental impact. So the coopetitive strategy, in our model, consists in implementing a set of policy decisions whose purpose is to be environmental sustainable and to enforce the global green economy. This is why the coopetitive variable in our model is represented by a vector variable whose components guarantee the achievement of the environmental sustainability of a global green economy. Thus, this original ES coopetitive model, applied at the global green economy, aims to enrich the set of tools for environmental policies.

\subsubsection{Win-win solutions.} The model provides several possible win-win solutions, strategic situations in which each country takes advantages by cooperating and competing at the same time, more than simply competing without cooperating; this shows the convenience for each country to participate actively to an environmental sustainability program within a coopetitive framework. The  model is based on a game theory framework that  provides a set of possible solutions in a coopetitive context, allowing us to find bargaining Pareto solutions in a win-win scenario (see also \cite{Ca-Sch TPREF}).

\subsubsection{Strategies.} The coopetitive strategies in our ES coopetitive model are:

\begin{enumerate}

\item investment  in maintenance of natural renewable resources;

\item investment in \emph{green technologies} against pollution (air, water);

\item incentives and taxes to change the patterns of consumption of the households.

\end{enumerate}

\section{The model: an ES coopetitive model of Economy}

The coopetitive model we propose hereunder must be interpreted as a normative model, in the sense that it will show the more appropriate solutions and win-win strategies chosen within a cooperative perspective, under precise conditions imposed by assumption.

\subsection{Strategies}

The strategy sets of the model are:

\begin{enumerate}

\item the set $E$ of strategies $x$ of a certain country $c$ - the possible aggregate biological-food production of the country $c$ - which directly influence both payoff functions, in a proper game theoretic approach \emph{\'a la Cournot};

\item the set $F$ of strategies $y$ of the rest of the word $w$ - the possible aggregate biological-food production of the rest of the world $w$ - which influence both pay-off functions; 

\item the set $C$ of $3$-dimensional shared strategies $z$, set which is determined together by the two game players, $c$ and the rest of the world $w$.

\end{enumerate}

\subsubsection{Interpretation of the cooperative strategy.} Any vector $z$ in $C$ is the $3$-level of aggregate investment for the ES economic approach, specifically $z$ is a triple $(z_1,z_2,z_3)$, where:

\begin{enumerate}

\item the first component $z_1$ is the aggregate investment, of the country $c$ and of the rest of the world $w$, in \textit{maintenance of natural renewable resources};

\item the second component $z_2$ is the aggregate investment, of the country $c$ and of the rest of the world $w$, in \textit{green technologies against pollution};

\item the third component $z_3$ is the aggregate algebraic sum, of the country $c$ and of the rest of the world $w$, of \textit{incentives (negative) and taxes (positive) to change the patterns of consumption of the households}.

\end{enumerate}

In the model, we assume that $c$ and $w$ define ex-ante and together the set $C$ of all cooperative strategies and (after a deep study of the coopetitive interaction) the triple $z$ to implement as a possible component solution.

\subsection{Main strategic assumptions}

We assume that any real number $x$, in the canonical unit interval $E := \IU = [0,1]$, is a possible level of aggregate production of the country $c$ and any real number $y$, in the same unit interval $F := \IU$, is the analogous aggregate production of the rest of the world $w$.

\subsubsection{Measure units of the individual strategy sets.} We assume that the measure units of the two intervals $E$ and $F$ be different: the real unit $1$ in $E$ represents the maximum possible aggregate production of country $c$ of a certain biological product and the real unit $1$ in $F$ is the maximum possible aggregate production of the rest of the word $w$, of the same good. Obviously these two units represents totally different quantities, but - from a mathematical point of view - we need only a rescale on $E$ and a rescale on $F$ to translate our results in real unit of productions.

\subsubsection{Cooperative strategy.} Moreover, a real triple (3-vector) $z$, belonging to the canonical cube $C: = \IU^3$, is the 3-investment of the country $c$ and of the rest of the world $w$ for new low-carbon innovative technologies, in the direction of sustainability of natural resources and for the environmental protection. Also in this case, the real unit $1$ of each factor of $C$ is, respectively:

\begin{enumerate}

\item the maximum possible aggregate investment  in maintenance of natural renewable resources;

\item the maximum possible aggregate investment in Ògreen technologiesÓ against pollution (air, water);

\item the maximum possible aggregate algebraic sum of incentives and taxes to change the patterns of consumption of the households.

\end{enumerate}

Let us assume, so, that the country and the rest of the world decide together, at the end of the analysis of the game, to contribute by a 3-investment $z = (z_1,z_2,z_3)$. We also consider, as payoff functions of the interaction between the country $c$ and the rest of the word $w$, two \emph{Cournot type} payoff functions, as it is shown in what follows.

\subsection{Payoff function of country $c$}

We assume that the payoff function of the country $c$ is the function $f_1$ of the unit 5-cube $\IU^5$ into the real line, defined by 
$$f_1(x, y, z) = 4x (1 - x - y) + m_1z_1 + m_2z_2 + m_3z_3 = 4x (1 - x - y) + (m|z),$$
for every triple $(x, y, z)$ in the 5-cube $\IU^5$, where $m$ is a characteristic positive real 3-vector representing the marginal benefits of the investments decided by country $c$ and by the rest of the world $w$ upon the economic performances of the country $c$.

\subsection{Payoff function of the rest of the world $w$}

We assume that the payoff function of the rest of the world $w$, in the examined strategic interaction, is the function $f_2$ of the 5-cube $\IU^5$ into the real line, defined by
$$f_2(x, y, z) = 4y (1 - x - y) + (n|z),$$
for every triple $(x, y, z)$ in the 5-cube $\IU^5$ , where $n$ is a characteristic positive real 3-vector representing representing the marginal benefits of the investments decided by country $c$ and the rest of the world $w$ upon the economic performances of the rest of the world $w$ itself. Note the symmetry in the influence of the pair $(m,n)$ upon the pair of payoff functions $(f_1,f_2)$.

\subsection{Payoff function of the coopetitive game}

We have so build up a coopetitive gain game $G = (f, \ge)$, with payoff function $f : \IU^5 \to \IR^2$, given by 
\begin{eqnarray*}
f(x, y, z) & = & (4x (1 - x - y) + (m|z), 4y (1 - x - y) + (n|z)) = \\
	   & = & 4(x (1 - x - y), y (1 - x - y)) + \sum z(m:n),
\end{eqnarray*}
for every triple $(x, y, z)$ in the compact 5-cube $\IU^5$, where $(m:n)$ is the 3-family of 2-vectors
$((m_i, n_i))^3_{i=1}$ (that is the family $((m_1, n_1), (m_2, n_2), (m_3, n_3))$) and where $\sum z(m : n)$ denotes the linear combination $\sum ^3_{i=1} z_i (m_i, n_i)$, of the family $(m:n)$ by the system of coefficients $z$.

\section{Study of the game $G = (p, \ge)$}

Fixed a cooperative strategy $z$ in the cube $\IU^3$, the game $G(z) = (p_z, \ge)$, with payoff function $p_z$ defined on the square $\IU^2$ by $p_z(x, y) = f(x, y, z),$ is the translation of the game $G(0_3)$ by the vector $v(z) = \sum z(m : n)$,
so that we can study the game $G(0_3)$ ($0_3$ is the origin of the Euclidean 3-space) and we can translate the various information of the game $G(0_3)$ by the vector $v(z)$. So, let us consider the game $G(0_3)$. This game $G(0_3)$ has been studied by Carf\`{\i} and Perrone in \cite{Carf-Perrone} (see also \cite{Carf2009}, \cite{Ca-Ri 2010} and \cite{Ca-Ri 2012}). The payoff conservative part (part of the payoff space greater than the conservative bi-value $(0,0)$) is the canonical 2-simplex $\IT$, convex envelope of the origin and of the canonical basis $e$ of the Euclidean plane $\IR^2$. This conservative part is represented in fig.1.

\begin{figure}[htbp]
\begin{center}
\includegraphics [width=4cm]{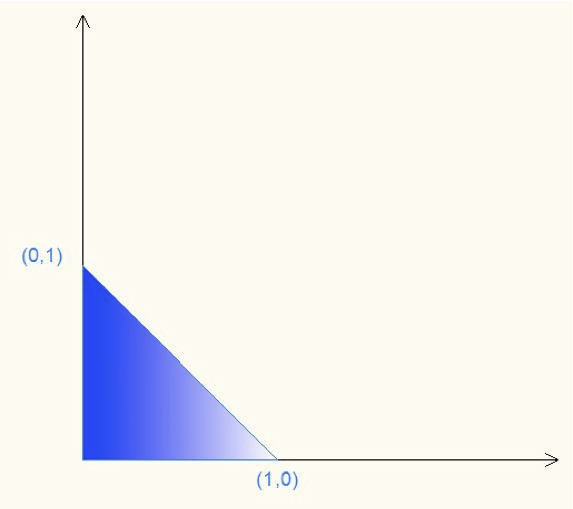} 
\caption{The conservative part of the Cournot payoff space, i.e. the positive part of the image $p_0 (\IU^2)$.}
\label{default}
\end{center}
\end{figure}

\subsubsection{Dynamical interpretation of coopetition.} In what follows we are interested in the trajectory of the dynamic path generated by the above conservative part $\IT$ in the co-opetitive evolution determined by the function $f$. The multi-time dynamic path we are interested in is the set-valued function
$\gamma : \IU^3 \to \IR^2 : z \mapsto \Gamma(z),$
where $\Gamma(z)$ is the conservative part of the game $G(z)$. The trajectory is nothing but the union of all configurations determined by the path $\gamma$: what we shall call our coopetitive payoff space. 

\subsection{Payoff space and Pareto boundary of the payoff space of $G(z)$}

The Pareto boundary of the payoff space of the $z$-section normal-form game $G(z)$ is the segment $[e_1, e_2]$, with end points the two canonical vectors of the Cartesian vector plane $\IR^2$, translated by the vector
$v(z) = \sum z(m : n),$ this is true for all 3-strategy $z$ in the unit cube $\IU^3$ (3-dimensional set).

\subsection{Payoff space of the co-opetitive game $G$}

The payoff space of the co-opetitive game $G$, the image of the payoff function $f$, is the union of the family of payoff spaces
$(p_z(\IU^2))_{z\in C},$ that is the convex envelope of the of points 0, $e_1$, $e_2$, and of their translations by the three vectors: $v(1,0,0) = (m_1, n_1)$; $v(1,1,0) = \sum^2_{i=1}(m_i, n_i)$ and $v(1,1,1) = \sum^3_{i=1}(m_i, n_i).$
 
\subsubsection{The construction of the Pareto maximal coopetitive path}

We show, in the following five figures, the construction of the coopetitive payoff space in three steps, in the particular case in which $m = (-1, 1, 1)$ and $n = (2, 1, -1)$, just to clarify the procedure. Moreover we shall consider here only the coopetitive space $S$ generated by the Pareto maximal boundary $\IM_2 = [e_1, e_2]$, since the Pareto Maximal boundary of the coopetitive game $G$ is contained in this part $S$. 

\begin{figure}
\begin{center}
\subfigure[$\IM_2$.\label{step0}] 
{\includegraphics[width=0.45\textwidth]{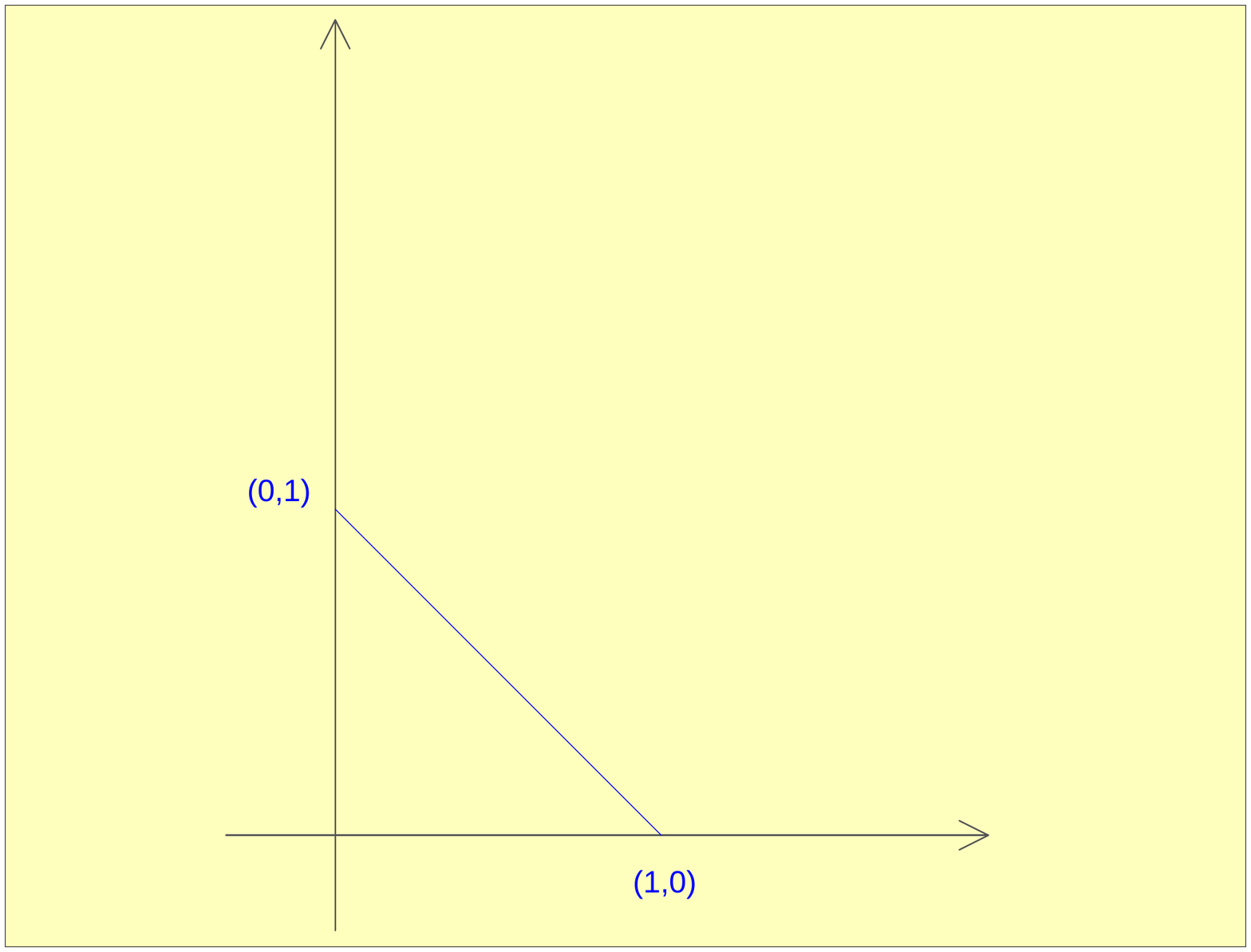}}
\subfigure[$\IM_2 + \IU(-1, 2)$.\label{step1}]
{\includegraphics[width=0.45\textwidth]{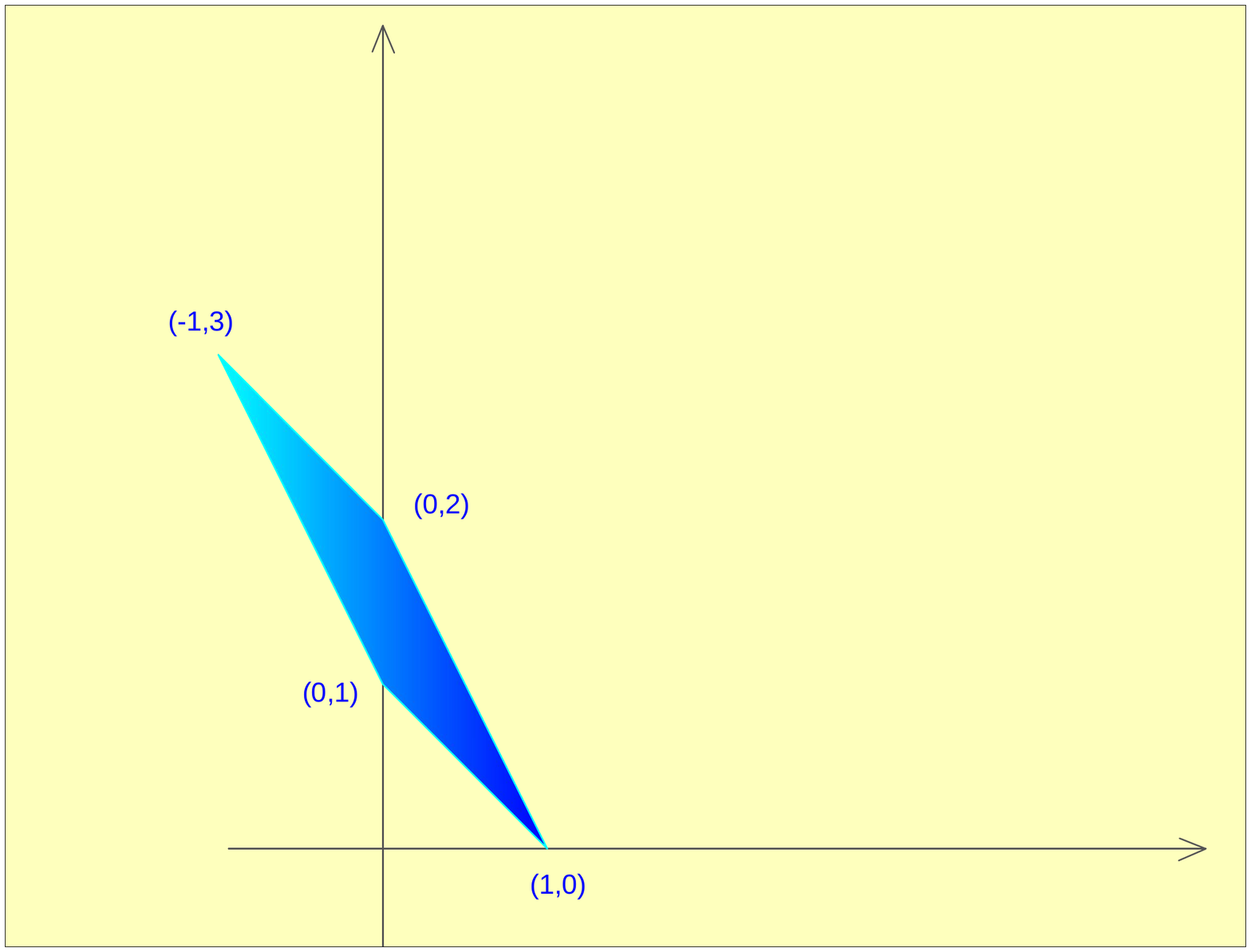}}
\caption{Step 0 and step 1}
\label{step0and1}
\end{center}
\end{figure}




\begin{figure}[htbp]
\begin{center}
\includegraphics [width=9cm]{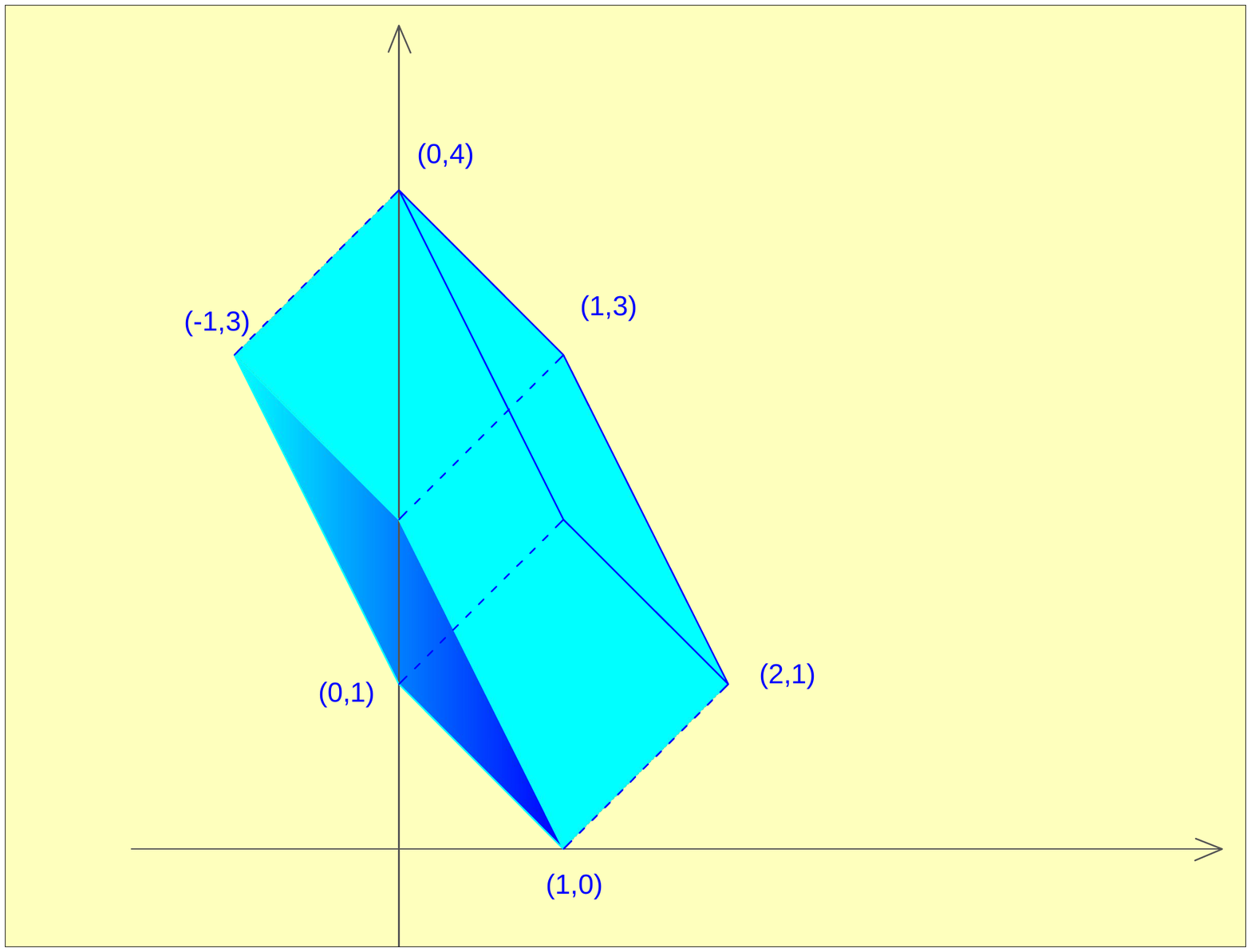} 
\caption{Second step: $\IM_2 + \IU(-1, 2) + \IU(1,1)$.}
\label{default}
\end{center}
\end{figure}

\begin{figure}[htbp]
\begin{center}
\includegraphics [width=10cm]{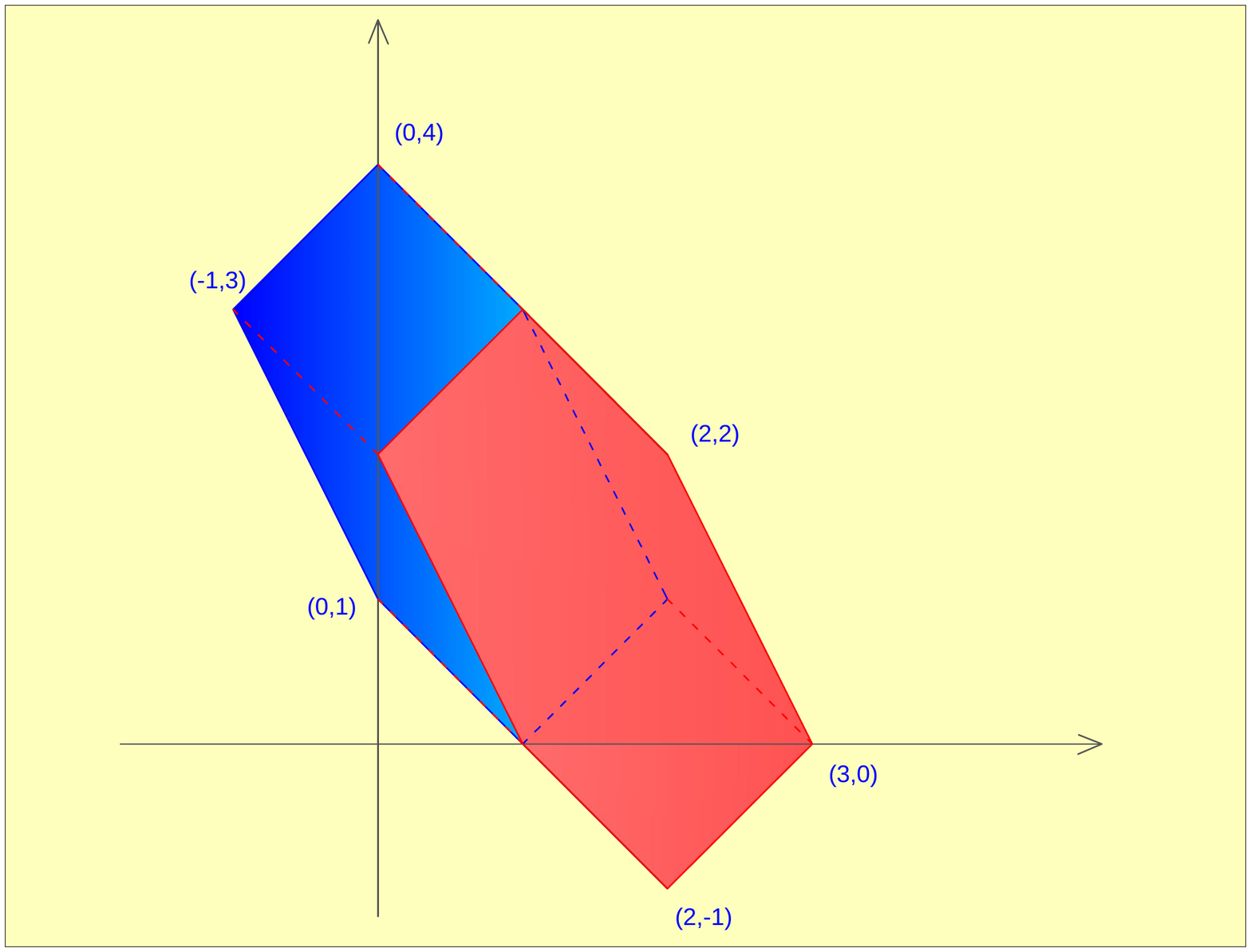} 
\caption{Third and final step: $\IM_2 + \IU(-1,2) + \IU(1,1) + \IU(1,-1)$.}
\label{default}
\end{center}
\end{figure}

\begin{figure}[htbp]
\begin{center}
\includegraphics [width=10cm]{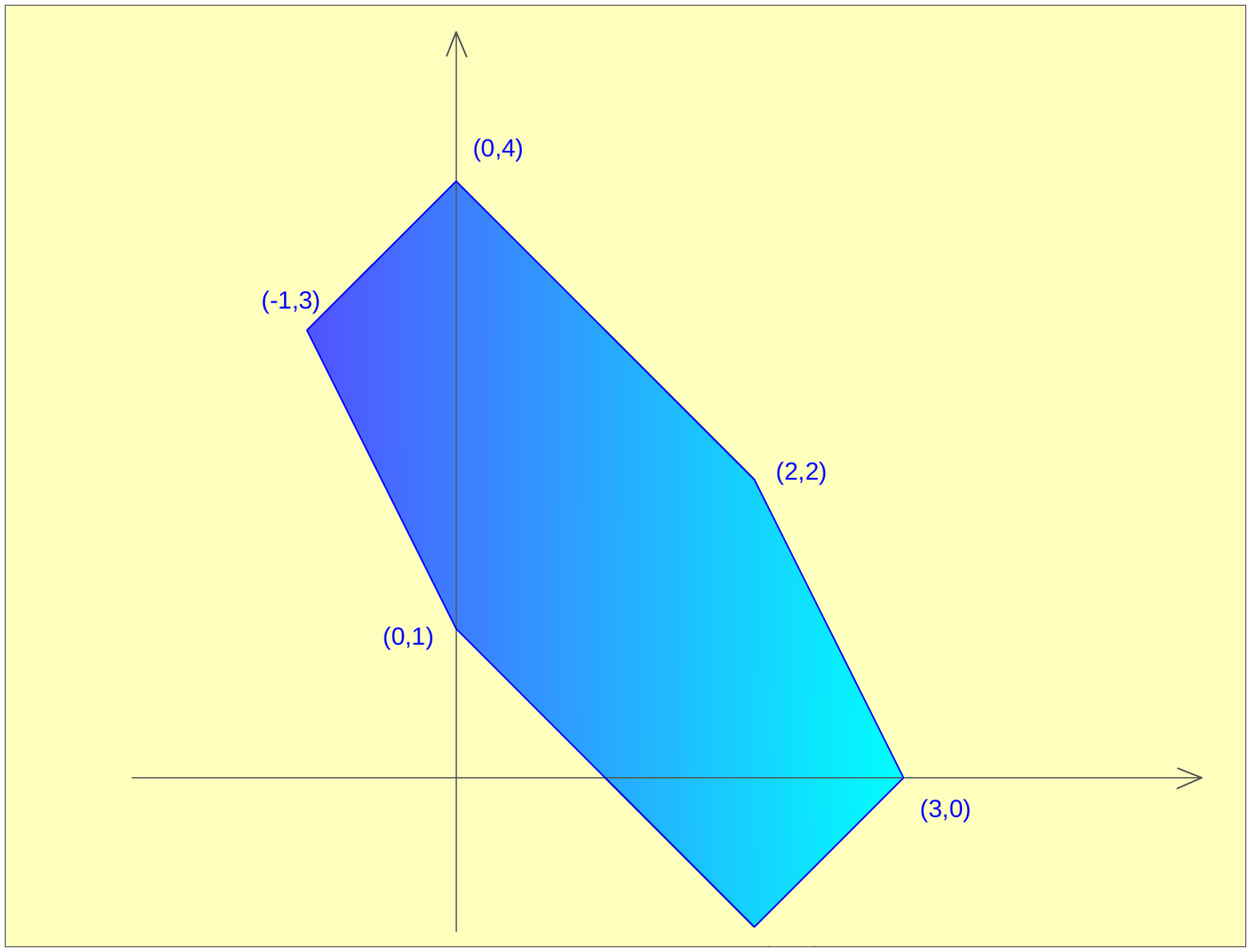} 
\caption{The coopetitive dynamical path of the initial Pareto boundary $\IM_2$.}
\label{default}
\end{center}
\end{figure}

\subsubsection{The Pareto maximal boundary of the payoff space $f(\IU^5)$.} The Pareto maximal boundary of the payoff space $f(\IU^5)$ of the coopetitive game $G$ is the union of segments
$[P', Q']\cup[Q',R'],$ where the point $P'$ is $(0,4)$, the point $Q'$ is $(2,2)$ and the point $R'$ is $(3,0)$.

\section{Solutions of the model}

\subsection{Properly coopetitive solutions}

In a purely coopetitive fashion, the solution of the coopetitive game $G$ must be searched for in the coopetitive dynamic evolution path of the Nash payoff $N' = (4/9, 4/9)$. Let us study this coopetitive dynamical path. We have to start from the Nash payoff $N'$ and then generate its coopetitive trajectory
$\mathcal{N} := N' + \IU(-1,2) + \IU(1,1) + \IU(1,-1).$
A \textit{purely coopetitive} solution is obtainable by cooperating on the set $C$ and (interacting) competing \emph{\'a la Nash} in a suitable best compromise game $G(\bar{z})$, where the cooperative strategy $\bar{z}$ is suitably and cooperatively chosen by the two players, by solving a bargaining problem. To be more precise, we give the following definition of properly coopetitive solution.

\textbf{Definition (of purely coopetitive solution.} \emph{We say that a triple $t = (x,y,z)$ is a properly coopetitive solution of a coopetitive game $G = (f, \ge)$ if its image $f(t)$ belongs to the maximal Pareto boundary of the payoff Nash Path of $G$.}

\subsubsection{Interpretation.} The complex game just played was developed with reference to the strategic triple $(x, y, z)$, and the functional relation $f$, which represents a continuous infinite family of games \emph{\'a la Cournot}, where in each member-game of the family the quantities are the strategies which vary in order to establish the Cournot-Nash equilibrium, and where the vector $z$ allows the game to identify possible cooperative solutions in a competitive environment, thus we have obtained a \emph{pure co-opetitive solution}.

\subsection{Construction of the Nash path}

As before, we can proceed step by step, as the following 5 figures just show.

\begin{figure}
\begin{center}
\subfigure[$N'$.\label{step2}] 
{\includegraphics[width=0.45\textwidth]{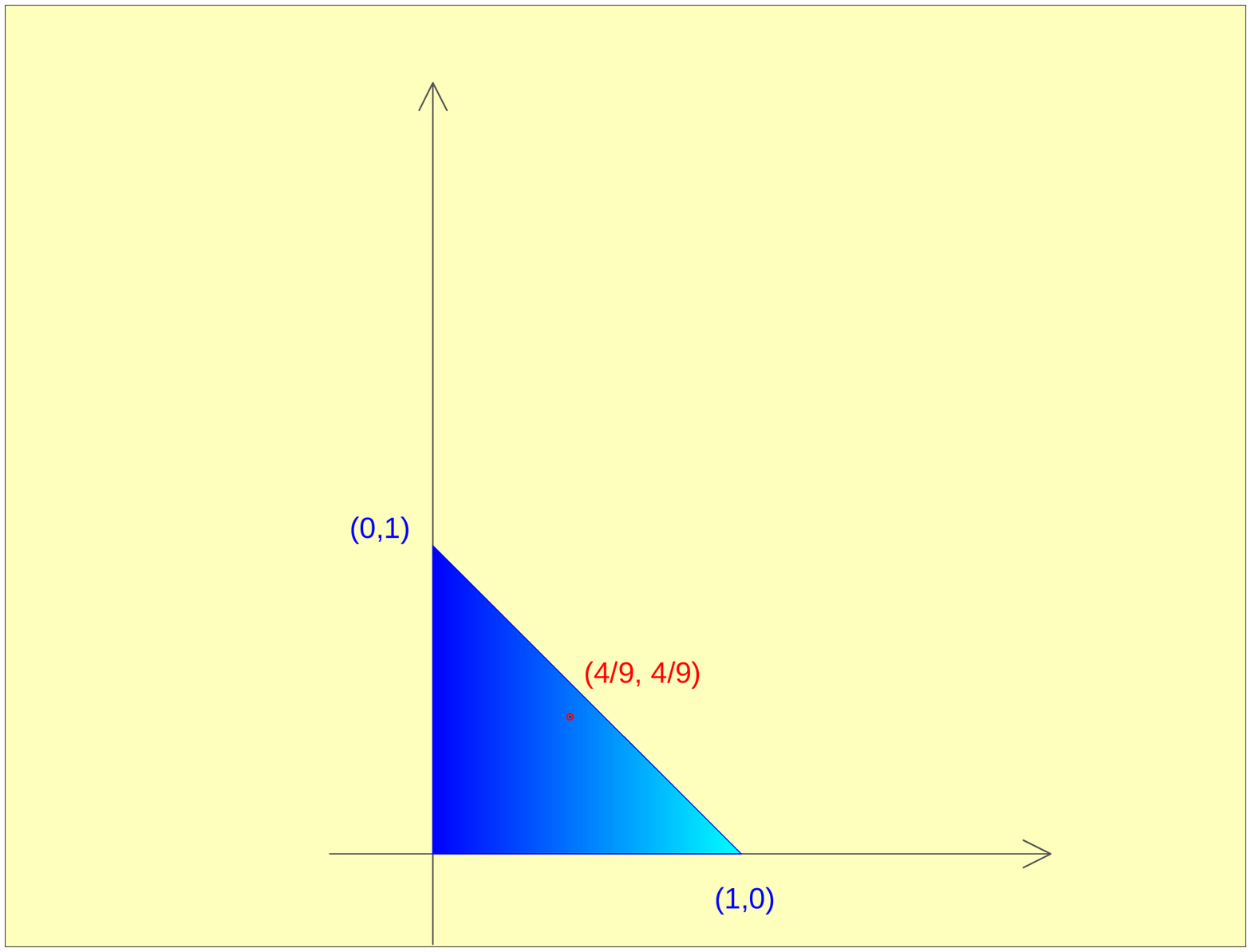}}
\subfigure[$N' + \IU(-1, 2)$.\label{step3}]
{\includegraphics[width=0.45\textwidth]{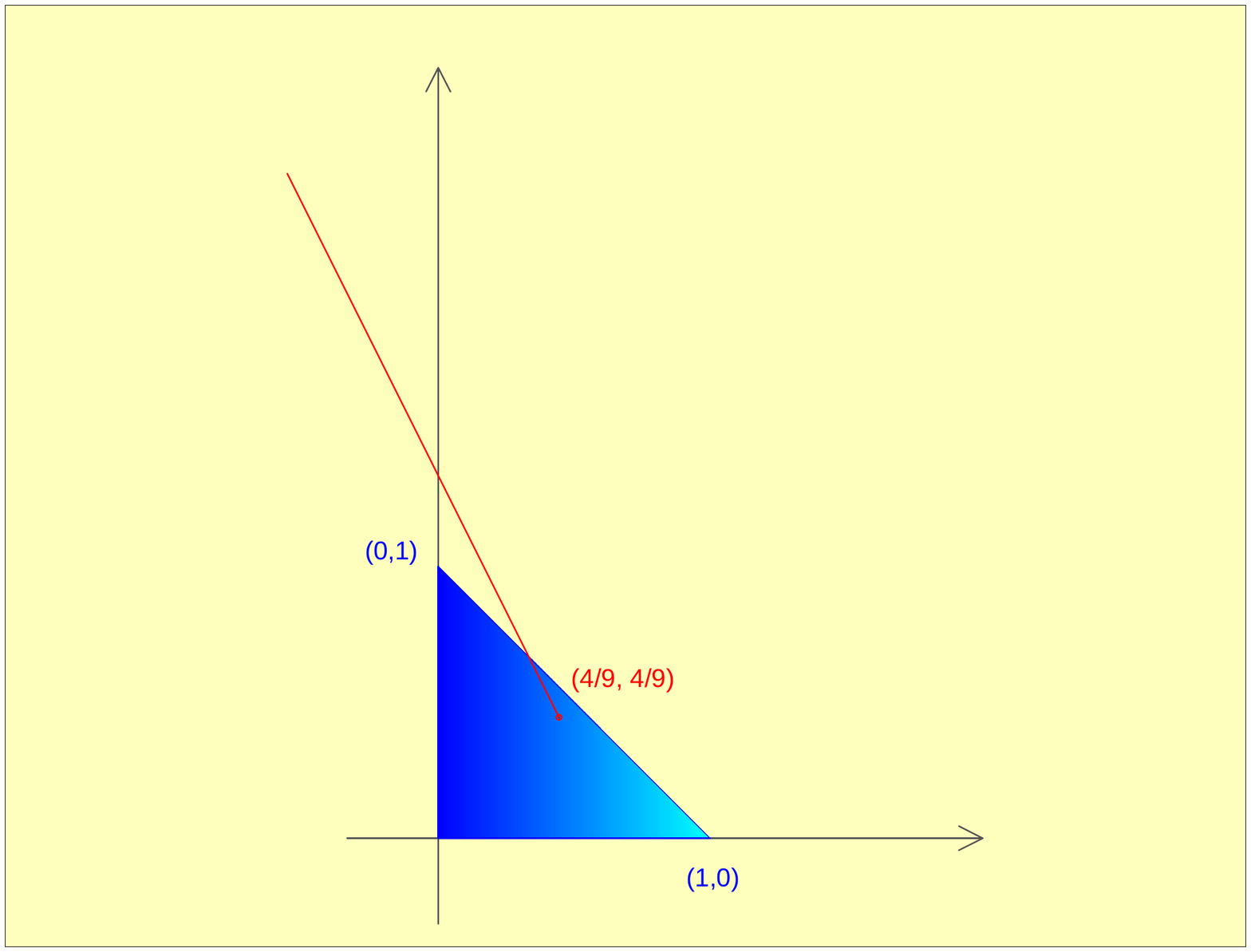}}
\caption{Step 0 and step 1}
\label{step0and1}
\end{center}
\end{figure}



\begin{figure}
\begin{center}
\subfigure[$N' + \IU(-1, 2) + \IU(1,1)$.\label{step2}] 
{\includegraphics[width=0.45\textwidth]{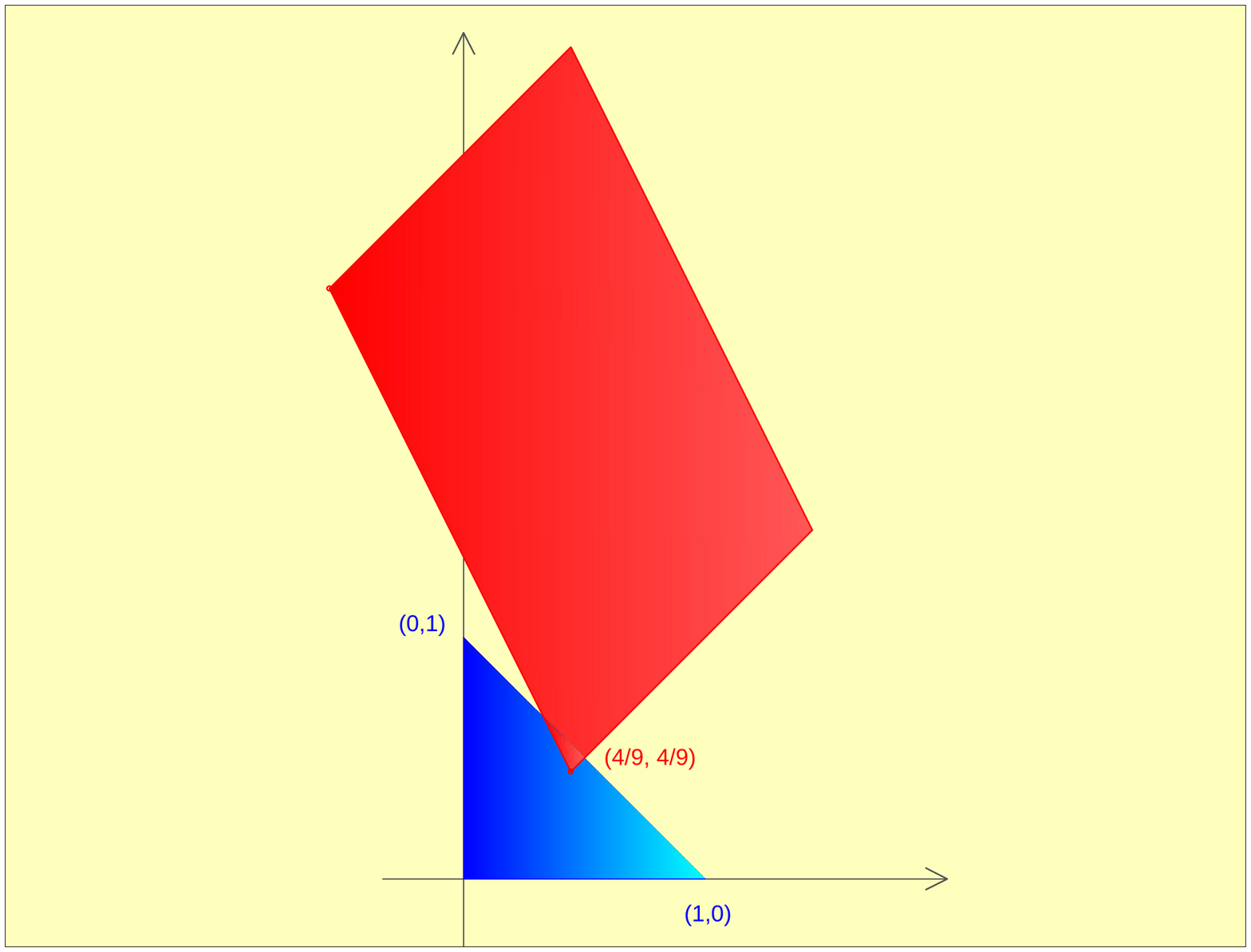}}
\subfigure[$N' + \IU(-1,2) + \IU(1,1) + \IU(1,-1)$.\label{step3}]
{\includegraphics[width=0.45\textwidth]{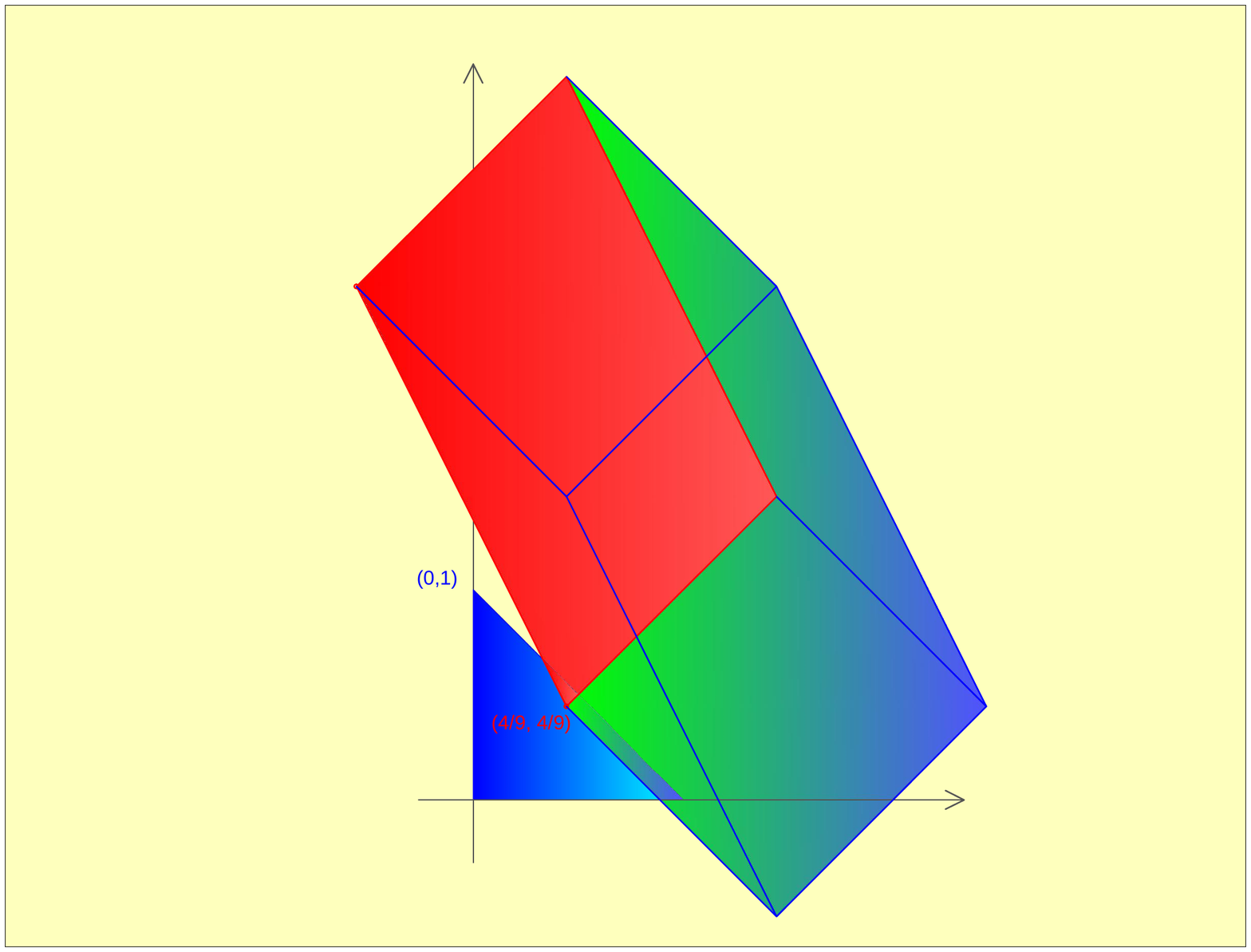}}
\caption{Step 2 and step 3}
\label{step0and1}
\end{center}
\end{figure}



\begin{figure}[htbp]
\begin{center}
\includegraphics [width=11cm]{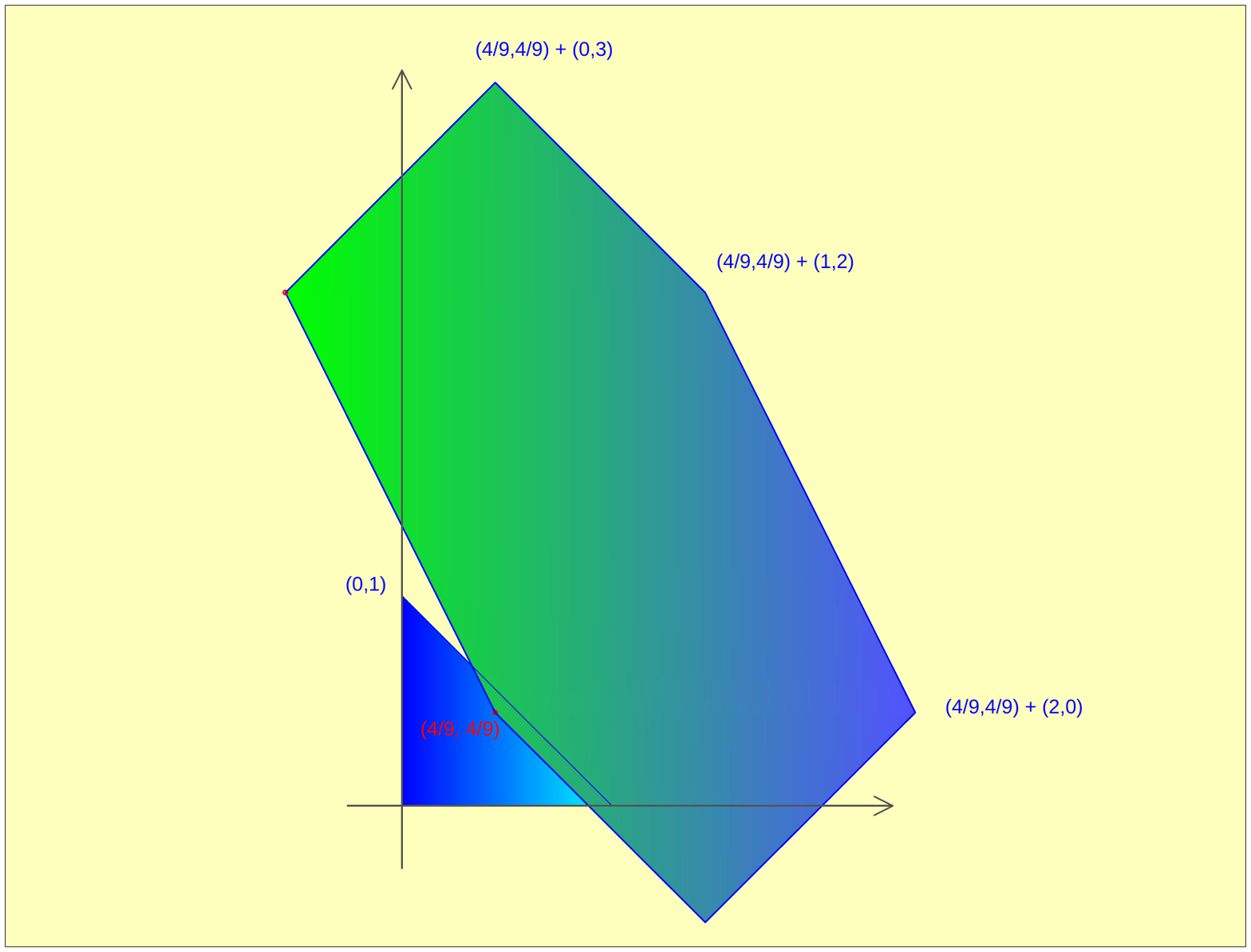}
\caption{The coopetitive dynamical path of the initial Nash equilibrium $N'$.}
\label{default}
\end{center}
\end{figure}

\subsubsection{Kalai-Smorodinsky purely coopetitive payoff solution.} The Kalai-Smorodinsky purely coopetitive payoff solution, with respect to the Nash payoff - the point $H$ in the following figure - is the solution of the classic bargaining problem $( \partial^* \mathcal{N}, N')$, where $\partial^*\mathcal{N}$ is the Pareto maximal boundary of the Nash path $\mathcal{N}$ and the threat point of the problem is the old initial Nash-Cournot payoff $N'$. The payoff solution $H$ is obtained by the intersection of the part of the Nash Pareto boundary which stays over $N'$ (in this specific case, the whole of the Nash Pareto boundary) and the segment connecting the threat point $N'$ with the supremum of the above part of the Nash Pareto boundary.

\begin{figure}[htbp]
\begin{center}
\includegraphics [width=11.5cm]{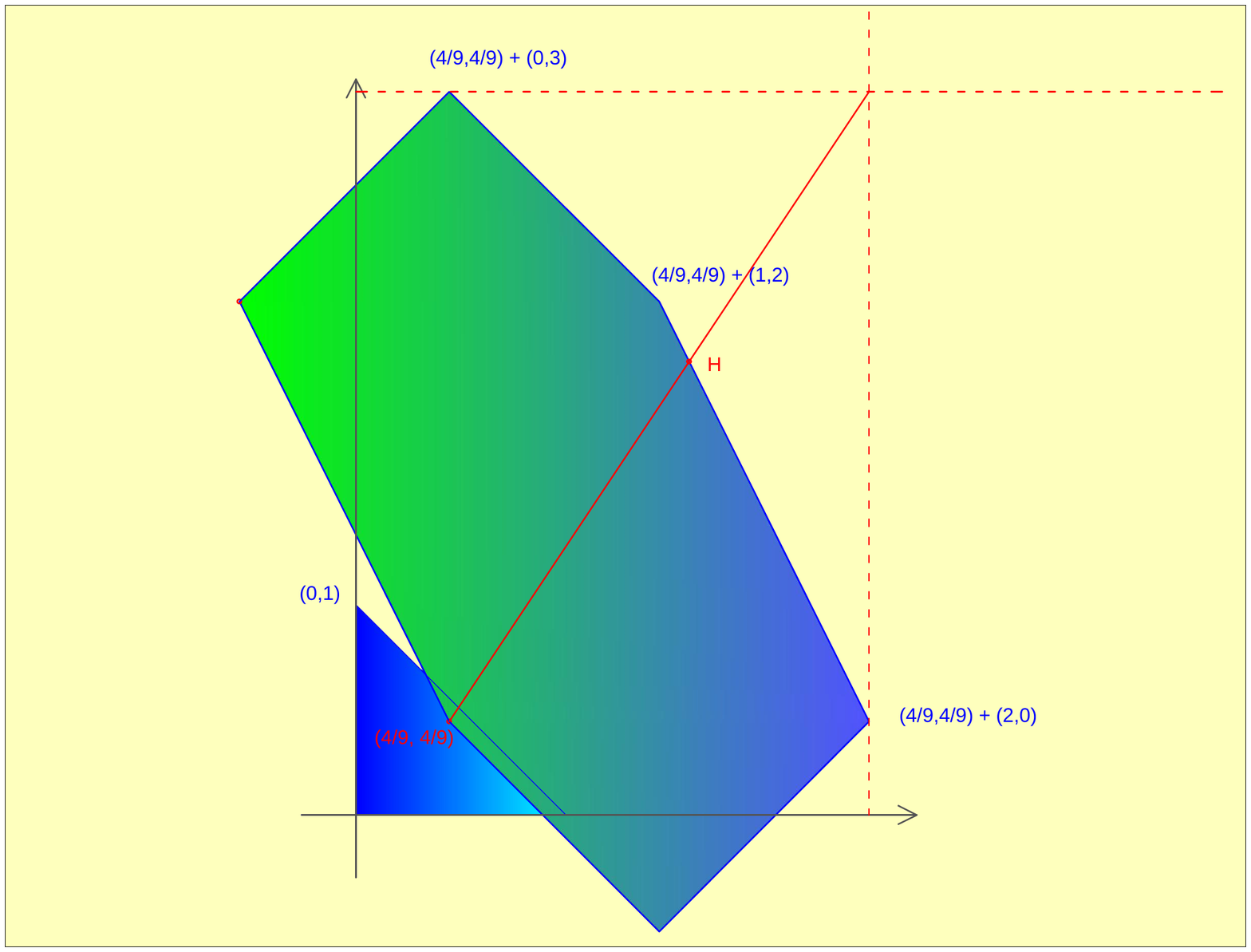} 
\caption{Kalai-Smorodinsky purely coopetitive payoff solution: $H$.}
\label{default}
\end{center}
\end{figure}

\subsubsection{Kalai-Smorodinsky purely coopetitive TU payoff solution.} In this game the Kalai-Smorodinsky purely coopetitive payoff solution is not optimal with respect to the Transferable Utility approach, indeed the TU Pareto boundary of our Nash path is the straight line
$$N' + (0,3) + \mathrm{span}(1,-1),$$
and the payoff $H$ does not belong to this line. The unique point of intersection among this TU boundary and the segment connecting the threat point $N'$ with the supremum of the Nash Pareto boundary is what we define the \textit{Kalai-Smorodinsky purely coopetitive TU payoff solution}.

\subsection{Super-cooperative solutions}

We can go further, finding a Pareto solution obtainable by double cooperation, in the following sense: we assume that in the game the two players will cooperate both on the cooperative 3-strategy $z$ and on the bi-strategy pair $(x,y)$.

\subsubsection{Super cooperative Nash bargaining solution.} The super cooperative Nash bargaining payoff solution, with respect to the infimum of the Pareto boundary, is by definition the point of the Pareto maximal boundary $M$ obtained by maximizing the real functional
$h : \IR^2 \to \IR$ defined by $(X,Y) \mapsto (X - \alpha_1)(Y -\alpha_2),$
where $\alpha$ is the infimum of the Pareto maximal boundary. In our case $\alpha$ is the origin of the plane, so this solution coincide with the medium point $Q' = (2,2)$ of the segment $[P', (4,0)]$. This point $Q'$ represents a win-win solution with respect to the initial (shadow maximum) supremum $(1,1)$ of the pure Cournot game, since it is strongly greater than $(1,1)$.

\subsubsection{Super cooperative Kalai-Smorodinsky bargaining solution.} The Kalai-Smorodinsky bargaining solution, with respect to the infimum of the Pareto boundary, coincide with the intersection of the diagonal segment $[\inf M,\sup M]$ and the Pareto boundary $M$ itself: the point $K = (12/7,16/7) $, of the segment $[P', Q']$. This point $K$ also represents a good win-win solution with respect to the initial (shadow maximum) supremum $(1,1)$ of the pure Cournot game.

\begin{figure}[htbp]
\begin{center}
\includegraphics [width=13cm]{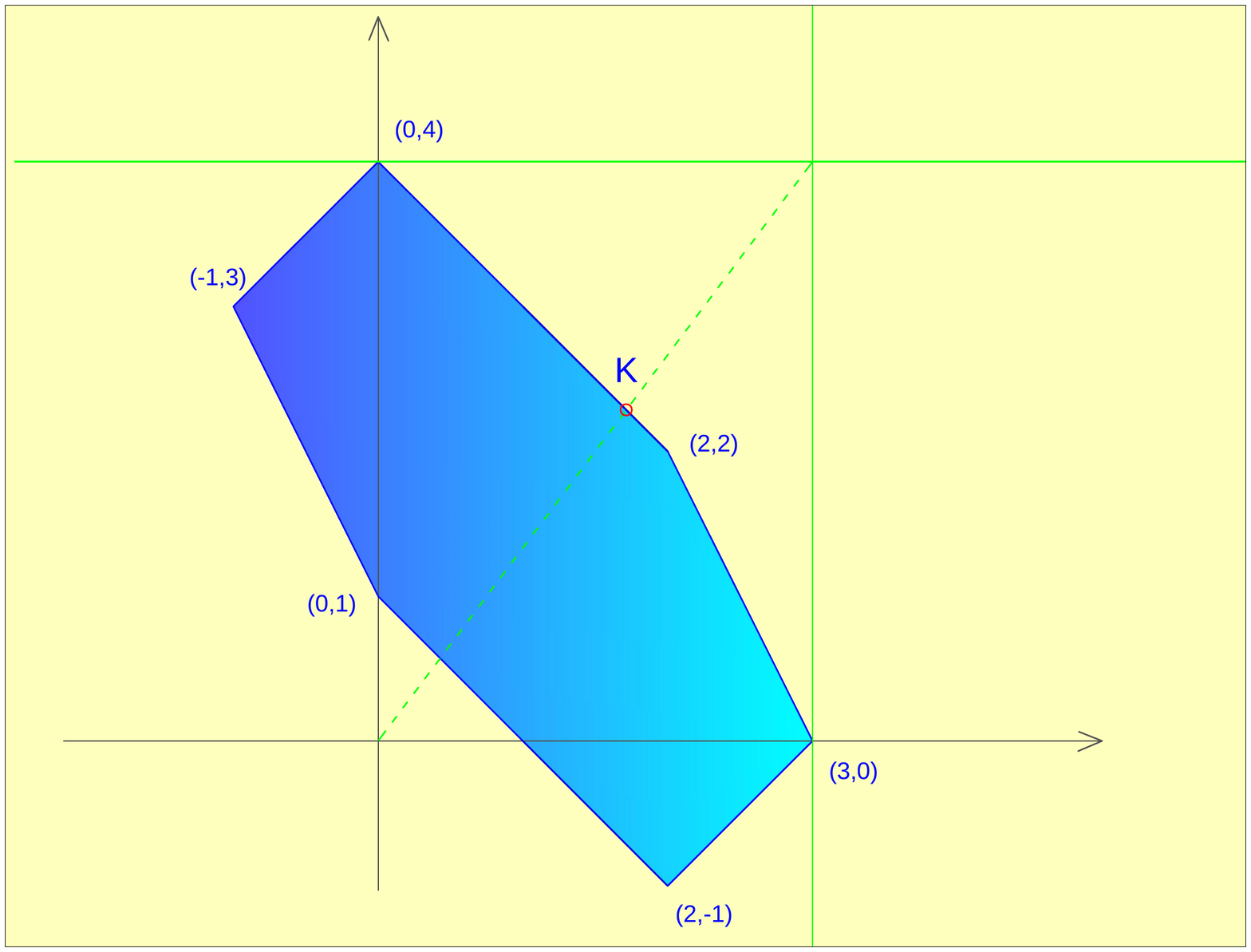} 
\caption{Super-cooperative Kalai-Smorodinsky solution in the payoff space: $K$.}
\label{default}
\end{center}
\end{figure}

\subsubsection{Super cooperative Kalai-Smorodinsky transferable utility solution.} The transferable utility solutions of our game in the payoff space are all the payoff pairs staying on the straight line
$\rm{aff}(P',Q') = (0,4) + \rm{span}(1,-1),$
the feasible TU solutions are those belonging to the segment $s = [(0,4),(3,1)]$. Note, anyway, that the Kalai-Smorodinsky solution of the bargaining problem $(s,0_2)$ is again the point $K$: the super cooperative Kalai-Smorodinsky transferable utility solution coincides with the super cooperative Kalai-Smorodinsky bargaining solution.

\subsection{Sunk costs}

For what concerns the sunk costs, we consider an initial bi-cost $(1, 1)$ necessary to begin the ES approach to the production, so that in a non-coopetitive environment we have a translation by the vector $(-1, -1)$ of the Nash equilibrium payoff $(4/9, 4/9).$
Although we have a bi-loss, in a co-opetitive environment the gain is strictly greater than the absolute value of the bi-loss, thus the new super-cooperative Kalai-Smorodinsky solution $K - (1,1)$ is greater than the old Nash equilibrium payoff.

\section{Conclusions} 

Our coopetitive model has tried to demonstrate which are the win-win solutions of a coopetitive strategic interaction that aims at a policy of Environment Sustainability and to implement a Green Economy. This policy concerns 

\begin{enumerate}
\item investment  in maintenance of natural renewable resources;

\item investment in green technologies against pollution (air, water);
\item incentives and taxes to change the patterns of consumption of the households;

\end{enumerate}
taking into account the sunk costs, and the determination of aggregate output of biological food of any country $c$ in a non-cooperative game \emph{\'a la Cournot} with the rest of the world.

The \emph{original analytical elements} that characterized our coopetitive model are the following:

\begin{enumerate}

\item firstly, we defined $z$ as the cooperative strategy, which is the instrumental 3-vector (3 dimensions) of the ES policy;

\item secondly, we adopted a non-cooperative game \emph{\'a la Cournot} for establishing an equilibrium bi-level $(x,y)$, that represents the levels of outputs of country $c$ and of the rest of the world $w$;

\item thirdly, we introduced the sunk costs of the ES approach;

\item finally, we suggested not only a pure coopetitive solution, but also two super-cooperative solutions on the coopetitive maximal Pareto boundary of our interaction, adopting the Nash bargaining and the Kalai-Smorodinsky methods, thus obtaining two best compromise solution.

\end{enumerate}

\subsubsection{Acknowledgements.} The authors wish to thank Dr. Eng. Alessia Donato for her valuable help in the preparation of the figures.



\end{document}